\documentclass[10pt]{article}                                                   
\usepackage{sectsty,latexsym,amssymb,dsfont,textcomp}                           
\usepackage[centertags]{amsmath}                                                
\usepackage{fullpage}                                                           
\usepackage[centertags]{amsmath}                                                
\sectionfont{\normalsize}                                                       
\subsectionfont{\normalsize}   
\title{The Schwinger model on $S^1$: Hamiltonian formulation, vacuum 
and anomaly}
\author{David Stuart
 \\{\it{\small Centre for Mathematical Sciences, Wilberforce Road, 
 Cambridge, CB3 OWA,
 England}}\\{\it\small{ email:dmas2@cam.ac.uk}}
}
\date{} 

\begin{document}
\baselineskip=10pt
\newcommand{\dirop}{\partial\hspace{-1.25ex}\slash}
\newcommand{\Dirop}{{D}\hspace{-1.55ex}\slash}
\newcommand{\bfG}{\mathbf{\Gamma}}
\newcommand{\bfU}{\mathbf{U}}
\newcommand{\bfH}{\mathbf{H}}
\newcommand{\bfPhi}{\mathbf{\Phi}}
\newcommand{\bfPi}{\mathbf{\Pi}}
\newcommand{\bfOm}{\mathbf{\Omega}}
\newcommand{\bfp}{\mathbf{\Psi_0}}
\newtheorem{lemma}{Lemma}
\newtheorem{prop}[lemma]{Proposition}
\newtheorem{theorem}[lemma]{Theorem}
\newtheorem{corollary}[lemma]{Corollary}
\newtheorem{definition}[lemma]{Definition}
\newtheorem{remark}[lemma]{Remark}
\newtheorem{Notation}[lemma]{Notation}
\newcommand{\sign}{\hbox{sign}\,}
\newcommand{\eps}{\epsilon}
\newcommand{\proof}{\noindent {\it Proof}\;\;\;}
\newcommand{\qed}{\protect~\protect\hfill $\Box$}
\newcommand{\id}{\mathds{1}}
\newcommand{\be}{\begin{equation}}
\newcommand{\ee}{\end{equation}}
\newcommand{\ba}{\begin{eqnarray}}
\newcommand{\ea}{\end{eqnarray}}
\newcommand{\bes}{\[}
\newcommand{\ees}{\]}
\newcommand{\bas}{\begin{eqnarray*}}
\newcommand{\eas}{\end{eqnarray*}}
\newcommand{\hoa}{{H^{1}_{\mbA}}}
\newcommand{\hta}{{H^{2}_{\mbA}}}
\newcommand{\linf}{{L^\infty}}
\newcommand{\tp}{{\tilde P}}
\newcommand{\cf}{{\cal F}}
\newcommand{\ch}{{\cal H}}
\newcommand{\cfnd}{{\cal F}^{n}-{\cal F}^{n-1}}
\newcommand{\p}{\partial}
\newcommand{\abs}[1]{\vert #1 \vert}

\newcommand{\spsi}{_{{{\Psi}}}}
\newcommand{\pt}{\frac{\partial}{\partial t}}
\newcommand{\pxk}{\frac{\partial}{\partial {x^k}}}
\renewcommand{\theequation}{\arabic{equation}}
\newcommand{\rgt}{\rightarrow}
\newcommand{\lngrgt}{\longrightarrow}
\newcommand{\intsT}{ \int_{0}^{T}\!\!\int_{\Sigma} }
\newcommand{\dxdt}{\;dx\,dt}
\newcommand{\sublt}{_{L^2}}
\newcommand{\sublf}{_{L^4}}
\newcommand{\naf}{\nabla_\mbA \Phi}
\newcommand{\covt}{(\pt -iA_0) }
\newcommand{\ano}{A^{n}_{0}}
\newcommand{\aNo}{A^{N}_{0}}
\newcommand{\Psino}{\Psi^{n}_{0}}
\newcommand{\PsiNo}{\Psi^{N}_{0}}
\newcommand{\Nmo}{{N\!-\!1}}
\newcommand{\nmo}{{n-1}}
\newcommand{\nmt}{{n-2}}
\newcommand{\Nmt}{{N\!-\!2}}
\newcommand{\gotm}{\frac{\gamma}{2\mu}}
\newcommand{\ootm}{\frac{1}{2\mu}}
\newcommand{\tloc}{T_{loc}}
\newcommand{\tmax}{T_{max}}
\font\msym=msbm10
\def\Real{{\mathop{\hbox{\msym \char '122}}}}
\font\smallmsym=msbm7
\def\smr{{\mathop{\hbox{\smallmsym \char '122}}}}
\def\Complex{{\mathop{\hbox{\msym\char'103}}}}
\newcommand{\wkarr}{\; \rightharpoonup \;}
\def\Weak{\,\,\relbar\joinrel\rightharpoonup\,\,}
\newcommand{\To}{\longrightarrow}
\newcommand{\rp}{\hbox{Re\,}}
\newcommand{\pa}{\partial_A}
\newcommand{\pbfa}{\partial_{\mathbf A}}
\newcommand{\pao}{\partial_{A_1}}
\newcommand{\pat}{\partial_{A_2}}
\newcommand{\dbar}{\bar{\partial}}
\newcommand{\barpa}{\bar{\partial}_{A}}
\newcommand{\barpbfa}{\bar{\partial}_{\mathbf A}}
\newcommand{\barpaphi}{\bar{\partial}_{A}\Phi}
\newcommand{\barpbfaphi}{\bar{\partial}_{\mathbf A}\Phi}
\newcommand{\myqed}{\hfill $\Box$}
\newcommand{\cd}{{\cal D}}
\newcommand{\dt}{\hbox{det}\,}
\newcommand{\sma}{_{{ A}}}
\newcommand{\bfpi}{{\mbox{\boldmath$\pi$}}}
\newcommand{\ce}{{\cal E}}
\newcommand{\ulh}{{\underline h}}
\newcommand{\ulg}{{\underline g}}
\newcommand{\ulX}{{\underline {\bf X}}}
\newcommand\la{\label}
\newcommand{\lamo}{\stackrel{\circ}{\lambda}}
\newcommand{\bfjo}{\underline{{\bf J}}}
\newcommand{\Vflato}{V^\flat_0}
\newcommand{\cm}{{\cal M}}
\newcommand{\dist}{{\mbox{dist}}}
\newcommand{\cs}{{\cal S}}
\newcommand{\mcH}{{\mathcal H}}
\newcommand{\mcD}{{\mathcal D}}
\newcommand{\mcK}{{\mathcal K}}
\newcommand{\mcF}{{\mathcal F}}
\newcommand{\mcn}{{\mathcal  V}}
\newcommand{\mce}{{\mathcal  E}}
\newcommand{\mcb}{{\mathcal B}}
\newcommand{\mca}{{\mathcal A}}
\newcommand{\mcdpsi}{{\mathcal D}_{\psi}}
\newcommand{\mcd}{{\mathcal D}}
\newcommand{\mcl}{{\mathcal L}}
\newcommand{\mclaphi}{{\mathcal L}_{(\mbA,\Phi)}}
\newcommand{\mcdadjp}{\mcdadj_{\psi}}
\newcommand{\mcdadj}{{\mathcal D}^{\ast}}
\newcommand{\zl}{{Z_\Lambda}}
\newcommand{\thl}{{\Theta_\Lambda}}
\newcommand{\ca}{{\cal A}}
\newcommand{\cb}{{\cal B}}
\newcommand{\cg}{{\cal G}}
\newcommand{\cu}{{\cal U}}
\newcommand{\co}{{\cal O}}
\newcommand{\smA}{\small A}
\newcommand{\ttheta}{\tilde\theta}
\newcommand{\tn}{{\tilde\|}}
\newcommand{\rbar}{\overline{r}}
\newcommand{\oeps}{\overline{\varepsilon}}
\newcommand{\cgl}{\hbox{Lie\,}{\cal G}}
\newcommand{\Ker}{\hbox{Ker\,}}
\newcommand{\Dom}{\hbox{Dom\,}}
\newcommand{\AC}{\hbox{AC\,}}
\newcommand{\const}{\hbox{const.\,}}
\newcommand{\Sym}{\hbox{Sym\,}}
\newcommand{\tr}{\hbox{tr\,}}
\newcommand{\grad}{\hbox{grad\,}}
\newcommand{\ttd}{{\tt d}}
\newcommand{\ttdel}{{\tt \delta}}
\newcommand{\ns}{\nabla_*}
\newcommand{\csl}{{\cal SL}}
\newcommand{\kr}{\hbox{Ker}}
\newcommand{\beq}{\begin{equation}}
\newcommand{\eeq}{\end{equation}}
\newcommand{\pr}{\hbox{proj\,}}
\newcommand{\proj}{{\mathbb P}}
\newcommand{\tproj}{\tilde{\mathbb P}}
\newcommand{\projq}{{\mathbb Q}}
\newcommand{\tprojq}{\tilde{\mathbb Q}}
\newcommand{\oN}{\overline N}
\newcommand{\cN}{\cal N}
\newcommand{\cmet}{{{\hbox{${\mathcal Met}$}}}}
\newcommand{\met}{{\hbox{Met}}}
\newcommand{\bfga}{{\mbox{\boldmath$\overline\gamma$}}}
\newcommand{\bfOmega}{\mbox{\boldmath$\Omega$}}
\newcommand{\bfTh}{\mbox{\boldmath$\Theta$}}
\newcommand{\bfmw}{{\bf m}_{\mbox{\boldmath$\smo$}}}
\newcommand{\bfmu}{{\mbox{\boldmath$\mu$}}}
\newcommand{\bfulX}{{\mbox{\boldmath${X}$}}}
\newcommand{\bfultX}{\tilde{\mbox{\boldmath${X}$}}}
\newcommand{\bfmuw}{{\mbox{\boldmath$\mu$}}_{\mbox{\boldmath$\smo$}}}
\newcommand{\dmug}{d\mu_g}
\newcommand{\ltwo}{{L^{2}}}
\newcommand{\lfour}{{L^{4}}}
\newcommand{\hone}{H^{1}}
\newcommand{\honea}{H^{1}_{\smA}}
\newcommand{\er}{e^{-2\rho}}
\newcommand{\onetwo}{\frac{1}{2}}
\newcommand{\lra}{\longrightarrow}
\newcommand{\dv}{\hbox{div\,}}
\newcommand{\mbA}{\mathbf A}
\newcommand{\mba}{\mathbf a}
\newcommand{\mbm}{\mathbf m}
\newcommand{\mbn}{\mathbf n}
\newcommand{\mbx}{\mathbf x}
\newcommand{\opsi}{\overline{\psi}}
\newcommand{\vac}{{\Omega_0}}
\def\smo{{\mbox{\tiny$\omega$}}}
\protect\renewcommand{\theequation}{\thesection.\arabic{equation}}

\font\msym=msbm10
\def\Real{{\mathop{\hbox{\msym \char '122}}}}
\def\R{\Real}
\def\Z{\mathbb Z}
\def\K{\mathbb K}
\def\J{\mathbb J}
\def\L{\mathbb L}
\def\D{\mathbb D}
\def\N{\mathbb N}
\def\Mink{{\mathop{\hbox{\msym \char '115}}}}
\def\Integers{{\mathop{\hbox{\msym \char '132}}}}
\def\Complex{{\mathop{\hbox{\msym\char'103}}}}
\def\C{\Complex}
\font\smallmsym=msbm7
\def\smr{{\mathop{\hbox{\smallmsym \char '122}}}}

\maketitle
\thispagestyle{empty}
\vspace{-0.3in}
\begin{abstract}
We present a Hamiltonian formulation of the
Schwinger model on the circle in Coulomb gauge
as a semi-bounded self-adjoint operator
which is invariant under the modular group ${\cal M}=\Z$ 
of large gauge transformations.  There is a nontrivial
action of $\cm$ on fermionic Fock space $\mcH_0$ 
and its vacuum which plays a role
analogous to that of the spectral flow in the formalism involving the infinite
Dirac sea. The formulation allows (i) a description
of the anomaly and its relation to this group action, and 
(ii) an explicit identification of the interacting vacuum
which arises after the destabilization of the non-interacting
vacuum in $\mcH_0$.
\end{abstract}

{MSC classification: 81T13}

{Keywords: Schwinger model, anomaly, gauge invariance, Hamiltonian construction}

\section{Introduction}
\setcounter{equation}{0}
\label{secint}
The Schwinger model is two dimensional quantum electrodynamics with
massless fermions. The action functional is 
\beq\la{act}
S=\int\, -\frac{1}{4}F_{\mu\nu}F^{\mu\nu}\,+\, 
\overline{\psi}i\Dirop_a\psi\;\;dx\,dt\,,\qquad
\Dirop_a=\gamma^\mu(\partial_\mu-{ie}a_\mu)
\eeq
describing the interaction of a Dirac spinor field
$\psi$ with an electromagnetic potential $a_\mu dx^\mu$ 
with associated 
electromagnetic field $F_{\mu\nu}=\partial_\mu a_\nu-\partial_\nu a_\mu$.
The model was shown to be formally solvable by Schwinger 
50 years ago in \cite{sch62},
and to possess the interesting property that although the classical field
theory describes massless particles, the quantum field theory describes
scalar bosons of mass $e/\sqrt{\pi}$. This feature is closely related to
the presence of an {\em anomaly}, i.e. a symmetry of the classical theory 
which is
not shared by the quantum theory (in this case chiral phase rotation
$\psi\mapsto e^{i\gamma^5\theta}\psi$).
A mathematical expression of the anomaly is that there is a 
current $j^{5,\mu}$ associated to the symmetry which is classically
conserved (i.e. divergence free) but in the quantum theory satisfies
\beq\la{ae}
\partial_\mu j^{5,\mu}=-\frac{1}{\pi}E,
\eeq
where
$E=\dot a-\partial a_0$ is the electric field.

In this letter we consider
the Schwinger model on the circle in Coulomb gauge, as in \cite{mant}, 
with particular attention paid to the role of the {\em modular} group $\cm$
of ``large'' gauge transformations which are left unfixed by Coulomb gauge.
But in contrast to the infinite Dirac sea formalism of \cite{mant} we 
use a positive energy representation (for the fermion field) which 
allows a formulation of the Hamiltonian  as a self-adjoint operator in a 
mathematically precise way, 
via classical bosonization results from \cite{liebmattis,uhl67}. The interest
lies in the rigorous and transparent explanation of the anomaly, and the related
appearance of the mass $e/\sqrt{\pi}$, in terms of a nontrivial action of $\cm$
on the non-interacting Fock vacuum - this latter action replaces
the role played by the spectral flow in the infinite Dirac sea formalism
for the anomaly given in \cite{mant}.

In this section we introduce as basis for the discussion
the expression for the second quantized Hamiltonian
(\eqref{finh}-\eqref{q5r}) which is derived from Schwinger
gauge invariant regularization in \cite{stu12}. 
As a starting point we take the Hamiltonian formulation of the
classical Schwinger model.
We work in $1+1$ dimensional
space-time with coordinates $(t,x)$ and
metric $dt^2-dx^2$, 
with $0\leq x\leq L$ and periodic boundary conditions. 
The dependent variables are a Dirac field $\psi(t,x)\in\C^2$
and an electromagnetic  connection form $a_\mu dx^\mu=a_0\,dt+a\,dx$.
The gauge transformations act as 
$$\psi\to e^{i{g}}\psi\qquad
\hbox{and}\qquad a_\mu\to a_\mu+\partial_\mu{g}
$$ 
where  ${g}={g}(t,x)$ 
is a sufficiently regular function {\em which is $L$ periodic in $x$}.
As in \cite{mant} we 
will work in the Coulomb gauge, in 
which the spatial component of the connection $a$
depends only on time so that the expression for the electric field
$E=\dot a-\partial a_0$ is in fact the decomposition into the
longitudinal and transverse components: $E^{long}=-\partial a_0$ and
$E^{tr}=\dot a$ respectively.
The time component $a_0$ is integrated out
via the Gauss law leading to the following classical Hamiltonian 
in the zero mass case:
\beq\la{cham}
\int_0^L
\frac{1}{2e^2}{\dot a}^2-\psi^\dagger\bigl(i\gamma^5(\partial-ia)\psi\bigr)
+\frac{1}{2}e^2(\psi^\dagger\psi) (-\Delta)^{-1}*(\psi^\dagger\psi)\;\; dx\,.
\eeq
Here $(-\Delta)^{-1}$ means the kernel of the operator $-\Delta=-\partial^2$
on $[0,L]$ with periodic boundary conditions, $*$ is convolution and
$\partial=\partial_x$.
Notice that the longitudinal component of the 
electric field
has been integrated out leaving only the 
transverse component $E^{tr}=\dot a$.
We use the following form of the
gamma matrices:
\beq\la{gam}
\gamma^0=\left(\begin{array}{cc}
1 & 0\\
0 & -1
\end{array}\right)\,,\quad
\gamma^1=\left(\begin{array}{cc}
0 & i\\
i & 0
\end{array}\right)\,,\quad
\gamma^5=\gamma^0\gamma^1=\left(\begin{array}{cc}
0 & i\\
-i & 0
\end{array}\right)\,,
\eeq
and we will use dots (resp. $\partial$) to indicate derivatives with
respect to $t$ (resp. $x$).
\begin{remark}\la{mgt}
Notice that, in contrast to the case when space is the whole real line,
the periodicity requirement means it is not possible 
to choose a gauge in which the spatial component of the connection $a$
is actually zero, only spatially constant.
However
in this formulation there is a residual gauge invariance by 
the {\em modular} group 
$$
\cm=\Z=\{g_N(x)=e^{2\pi iNx/L}\}_{N\in\Z}
$$ 
of {\em large}
gauge transformations. Notice as a first
consequence $a$ is now defined mod $2\pi/L$ so that it is now taking values
in the circle $S^1=\R/(2\pi/L)$ which is dual to the spatial domain
$\R/L$. We shall see that a careful treatment of the
residual component of the potential $a=a(t)$ and invariance under the group
$\cm$ illuminates greatly the
role of gauge invariance in producing the anomaly and 
the interacting vacuum.
\end{remark}
The classical equations of motion  associated to \eqref{cham} are
\begin{align}\begin{split}
i\dot\psi&=-i\gamma^5(\partial\psi-ia\psi)-a_0\psi\\
\dot E^{tr} &=\frac{e^2}{L}\,\int_0^L\,{\psi}^\dagger\gamma^5\psi
\,dx\,,\qquad \dot a=E^{tr}\,\,,\end{split}\la{ceom}
\end{align}
where $a_0$ is determined by the Gauss law constraint
$-\Delta a_0
={-e^2}\,{\psi}^\dagger\psi={-e^2}\,{j^0}\,$. We will write
$j^0={\psi}^\dagger\psi$ and $j^1={\psi}^\dagger\gamma^5\psi$
for the currents and $Q=\int_0^L j^0\,dx\,,Q^5=\int_0^L j^1\,dx\,$
for the corresponding charges. In order that the Gauss law admit
a periodic solution it is necessary that $Q=0$, so that {\em throughout
it will be assumed that the total charge is zero.} In the classical
theory both the electromagnetic current $j=j^{\mu}
=\overline\psi\gamma^\mu\psi=(j^0,j^1)$ and the
axial current $j^{5,\mu}=\overline\psi\gamma^\mu\gamma^5\psi=(j^1,j^0)$ are conserved, but in the quantum
theory only the first of these properties holds - the conservation
law for $j^5$ is replaced by the anomaly equation \eqref{ae}, 
see \S\ref{anom}. As a first intimation of the connection of \eqref{ae}
with mass generation notice that together with \eqref{ceom} it implies
that the electric field satisfies $(\Box+\frac{e^2}{\pi})E=0$
in place of $\Box E=0$ - thus the anomalous right hand side of
\eqref{ae} generates a mass $e/\sqrt{\pi}$. However rather than
start with \eqref{ae} we will approach
the problem through the Hamiltonian, and the
reason for the anomaly will appear in \S\ref{mod} as a consequence
of defining a regularized Hamiltonian which is invariant under the action of
$\cm$; see also the comments at the end of section
\ref{anom}, in which \eqref{ae} is finally derived.

To quantize the theory it is necessary to associate operators
to the fields which satisfy the canonical relations:
\beq\la{car}
\{\psi_\alpha(t,x),\psi_\beta^\dagger(t,y)\}=
\delta_{\alpha\beta}\delta(x-y)
\eeq
(other anti-commutators being zero),
and
\beq\la{ccr}
[E^{tr}, a]=[\dot a,a]=-\frac{ie^2}{L}
\eeq
(other commutators being zero). For the latter we will use the Schr\"odinger
representation in which $a$ is represented by coordinate multiplication
on $L^2([0,\frac{2\pi}{L}])$, while 
$$
E^{tr}=-\frac{ie^2}{L}\frac{d}{da}.
$$
In the absence of interaction with any matter fields the electromagnetic
field is described by the Hamiltonian $H_{em}=-\frac{e^2}{2L}\frac{d^2}{da^2}$
on $L^2([0,\frac{2\pi}{L}])$ with periodic boundary conditions - the large
gauge transformations described in remark \ref{mgt} are the reason that
periodic boundary conditions are appropriate. We shall see below how these
boundary conditions are modified in the presence of interactions with fermionic matter.

The relations \eqref{car} are interpreted in the positive energy representation
by writing
\beq\la{repn}
\psi=\frac{1}{\sqrt{L}}\sum_{n\in\Z} \bigl(b_n u_n e^{ik_nx}+c_n^\dagger v_n e^{-ik_n x}
\bigr)\,,\quad k_n=\frac{2n\pi}{L}
\eeq
with 
\beq\la{carm}
\{b_n,b_{n'}^\dagger\}=
\{c_n,c_{n'}^\dagger\}=
\delta_{nn'}
\eeq
(other anti-commutators being zero) and
\begin{align}\begin{split}
u_n&=u^R\id_{\{n\geq 0\}}+u^L\id_{\{n<0\}}\,,\\
v_n&=u^R\id_{\{n> 0\}}+u^L\id_{\{n\leq 0\}}
\,.\end{split}\la{g5ev}
\end{align}
The $u^{L,R}$ are eigenvectors of $\gamma^5$ with 
$\gamma^5 u^R=u^R$ and  $\gamma^5 u^L=-u^L$. 
The $b_m^\dagger\,,\, b_m$ (resp. $c_m^\dagger\,,\, c_m$) are fermionic
(resp. anti-fermionic) creation, annihlation operators
acting on the zero charge fermionic Fock space $\mcH_0$.
The total Hilbert space for the theory can now be defined as 
\beq\la{dhs}
\mcK=\{\Psi=\Psi(a)\in\mcH_0:\Psi\in L^2([0,\frac{2\pi}{L}];\mcH_0)
\}\,,
\eeq
with norm defined by $\|\Psi\|_{\mcK}^2=\int_0^{\frac{2\pi}{L}}\,\|\Psi\|^2\,da$
where $\|\,\cdot\,\|$ is the Fock space norm.
Recall the fermionic Fock space: 
there is a (non-interacting) vacuum ${\Omega_0}$ 
and associated finite particle states
\beq\la{defv}
\Omega_{\mbm,\mbn}=
\prod\, b_{m_i}^\dagger c_{n_j}^\dagger {\Omega_0}
\eeq
where $\mbm=\{m_i\}_{i=1}^M$ and $\mbn=\{n_j\}_{j=1}^N$
range over subsets of $\Z$ of arbitrary finite size. 
(In \eqref{defv} and similar formulae it will be assumed that
products of creation/annihlation operators are ordered from left to right
as the indices $m_i$ or $n_j$ increase. This fixes the overall sign.)
Let $\mcF$ be the linear
span of all the $\Omega_{\mbm,\mbn}$, let $\mcF_0\subset\mcF$ be the
zero charge subspace in which there are equal numbers of
fermions and anti-fermions, i.e. $M=N$. The zero charge Fock
space $\mcH_0$ 
is the completion of $\mcF_0$ in the Fock space norm $\|\,\cdot\,\|$,
and the vectors in \eqref{defv} constitute an orthonormal basis.
There is a self-adjoint operator which extends the operator given 
on $\mcF_0$ by
$$
Q^{5}=
\sum_{n\geq 0}
{b}^\dagger_{n}b_{n}-
\sum_{n<0}
{b}^\dagger_{n}b_{n}-
\sum_{n> 0} c_n^\dagger c_{n}
+\sum_{n\leq 0}c_n^\dagger c_{n}\,.
$$
which will also be denoted $Q^5$; it will be referred
to as the {\em axial} (or chiral) charge operator. Define
$\mcF_0^P\subset\mcF_0=\{\Ker(Q^5-2P)\}\cap\mcF_0$.
The corresponding completions are
denoted $\mcH_0^P$, and are the orthogonal eigenspaces arising
in the spectral decomposition of $Q^5$.

We define the {\em unexcited} states $\Omega_P$ in the case
when $M=N=P\in\Z^+$ as follows. The case $P=0$ corresponds to the 
vacuum ${\Omega_0}$.
For $P\geq 0$
let $m_i=n_i=i$ for $0\leq i\leq P$, and define
the unexcited state
\beq
\Omega_{P+1}\;=\;
\prod\limits_{i=0}^P\;\prod\limits_{j=0}^P\, b_{m_i}^\dagger 
c_{-n_j}^\dagger {\Omega_0}\qquad (P\in\Z^+)\,;
\eeq
for $P<0$ and  $M=N=-P$ let $m_i=n_i=-i$ for $0<i\leq -P$, and define
the unexcited state as
\beq
{\Omega_P}\;=\;\prod\limits_{i=-1}^P\;\prod\limits_{j=-1}^P\, b_{m_i}^\dagger 
c_{-n_j}^\dagger {\Omega_0}\qquad 
(P\in\Z^{-}-\{0\})\,.
\eeq
The representation of \eqref{car} used in \eqref{repn} is a positive energy
representation in that the free Dirac Hamiltonian  
$H^0_D=\sum_{m\in\Z}\,|k_m|\,({b}^\dagger_{m}b_{m}+
c_m^\dagger c_{m})\,\geq 0\,.$

In the process of quantizing 
it is necessary to define carefully what is meant mathematically 
by the various formal expressions for bilinear quantities such 
as those for the axial charge
and the Hamiltonian itself, which involve products of what are at best
operator valued distributions. As emphasized in \cite{sch62} the definition
needs to be chosen carefully to ensure gauge invariance is maintained.
The least intrusive way of doing this seems
to be by Schwinger regularization (point-splitting), and the relevant
computations are presented in some notes available online (\cite{stu12}).
The endpoint of this is the following
formula for the regularized Hamiltonian: $H=H_0+:H_{coul}:\,$, where
\beq\la{finh}
H_0=-\frac{e^2}{2L}\frac{d^2}{da^2}+
\sum_{m\in\Z}\,|k_m|\,({b}^\dagger_{m}b_{m}+
c_m^\dagger c_{m})\,-\,\frac{a^2L}{2\pi}-\,a Q^{5,reg}\,,
\eeq
and
\beq\la{coul}
H_{coul}=\frac{e^2L}{2}\sum_{m\neq 0}\frac{1}{k_m^{2}}\jmath^0(-m)\jmath^0(m)
\eeq
is
the Coulomb energy, written in terms of the fourier modes of the current
operator 
$$j^0=
\sum\jmath^0(m)e^{ik_mx}\,.$$
In \eqref{finh} the symbol $Q^{5,reg}$ indicates
the regularized axial charge operator given by:
\beq\la{q5r}
Q^{5,reg}=
\sum_{n\geq 0}
{b}^\dagger_{n}b_{n}-
\sum_{n<0}
{b}^\dagger_{n}b_{n}-
\sum_{n> 0} c_n^\dagger c_{n}
+\sum_{n\leq 0}c_n^\dagger c_{n}
-\frac{aL}{\pi}-1\,.
\eeq
This expression is also derived from  Schwinger regularization in \cite{stu12},
where it is also shown that 
the corresponding expression for the regularized ordinary charge 
is in fact unchanged, i.e. 
$$Q=Q^{reg}=\sum_{n\in\Z}
\bigl({b}^\dagger_{n}b_{n}-
c_n^\dagger c_{n}\bigr)\,.
$$
Formulae closely related to \eqref{finh}-\eqref{q5r},
but in the infinite Dirac sea context,
can be found in \cite[\S 3]{mant}, where they are derived using
a gauge invariant heat kernel regularization to handle the arbitrarily
unbounded negative energies which arise in that formulation.

The aim in this letter is to recall how clasical bosonization results
can be used to make sense of the above expression for $H$ 
as a self-adjoint operator
on the Hilbert space $\mcK$, and thence to clarify the vacuum structure 
and the anomaly in a mathematically precise way.
These clarifications hinge upon an understanding of the gauge transformations,
so we first discuss the action of the group $\cm$ of large gauge 
transformations. This leads to the correct boundary condition \eqref{bc} 
which is
required to complete the mathematical formulation of the Schwinger model
in terms of the Hamiltonian \eqref{finh}-\eqref{q5r}.

\section{Action of $\cm$ and twisted periodicity}
\la{mod}
We define a unitary action of the group 
$\cm=\Z=\{g_N(x)=e^{2\pi iNx/L}\}_{N\in\Z}$
of large gauge transformations on $\mcH_0$. The formulae are 
best motivated
by comparison with the natural expressions in the infinite Dirac sea - see
\cite{mant,stu12}. There is 
a unitary operator $\bfG$, corresponding to the generator $g_1$, whose
action on the non-interacting vacuum state is
\beq\la{gt3}
{\bfG}\Omega_0\,=\Omega_{-1}=\,b_{-1}^\dagger c_1^\dagger\Omega_0\,.
\eeq
The action on Fock space is then determined by specifying the action
on the set of creation and annihlation operators, on which 
it acts as a modified shift operator:
\begin{align}\la{gt1}\begin{split}
b_n&\to {\bfG} b_n{\bfG^{-1}}\,=\,b_{n-1}\,,\; n\neq 0\,,\qquad  b_0\to {\bfG} b_0{\bfG^{-1}}\,=\,c_1^\dagger
\\
c_n&\to {\bfG} c_n{\bfG^{-1}}\,=\,c_{n+1}\,,\; n\neq 0\,,\qquad c_0\to {\bfG} c_0{\bfG^{-1}}\,=\,b_{-1}^\dagger
\end{split}\end{align}
with corresponding relations for the adjoints:
\begin{align}\la{gt2}\begin{split}
b^{\dagger}_n&\to {\bfG} b^{\dagger}_n{\bfG^{-1}}\,=\,b^{\dagger}_{n-1}\,,\; 
n\neq 0\,,\qquad  b_0^{\dagger}\to {\bfG} b^{\dagger}_0{\bfG^{-1}}\,=\,c_1
\\
c^{\dagger}_n&\to {\bfG} c^{\dagger}_n{\bfG^{-1}}\,=\,c^{\dagger}_{n+1}\,,\; 
n\neq 0\,,\qquad c^{\dagger}_0\to {\bfG} c^{\dagger}_0{\bfG^{-1}}\,=\,b_{-1}
\end{split}\end{align}
\begin{lemma}\la{ag}
The formulae \eqref{gt3}-\eqref{gt2} determine an action of $\cm=\Z$
on $\mcH_0$ generated by $\bfG$, with
the property that ${\bfG}{\Omega_P}=\Omega_{P-1}$ for all $P$.
Similarly there is a corresponding
modified shift action for the inverse $\bfG^{-1}$ 
with \eqref{gt3}-\eqref{gt2} inverted, so that in particular
$\bfG^{-1}\cdot{\Omega_0}=b_0^\dagger c_0^\dagger\Omega_0
=\Omega_{1}$ and more generally $\bfG^{-1}\cdot{\Omega_P}=\Omega_{P+1}$.
\end{lemma}
\proof
This is a consequence of the fact that $\bfG$ acts as a bijection on the
set of orthonormal basis vectors $\{\Omega_{\mbm,\mbn}\}$ labelled
by pairs of finite subsets of $\Z$: in fact
$\bfG \Omega_{\mbm,\mbn}=\iota_{\mbm,\mbn}\,\Omega_{\mbm',\mbn'}$ where 
$\iota_{\mbm,\mbn}\in\{\pm 1\}$ is an unimportant overall sign and
the subsets $\mbm',\mbn'$
are given by
\begin{align}
\mbm'&=\cup_{m\in\mbm,m\neq 0}\{m-1\}\;\;\qquad\qquad\quad (0\in\mbn)\\
&=\cup_{m\in\mbm,m\neq 0}\{m-1\}\cup\{-1\}\,\qquad (0\notin\mbn)\\
\mbn'&=\cup_{n\in\mbn,n\neq 0}\{n+1\}\;\;\qquad\qquad\qquad (0\in\mbm)\\
&=\cup_{n\in\mbn,n\neq 0}\{n+1\}\cup\{+1\}\,\quad\qquad (0\notin\mbm)\,.
\end{align}
This is clearly a bijection on 
pairs of subsets of $\Z$ of arbitrary finite size, whose
inverse is of the same form: $\bfG^{-1} \Omega_{\mbm,\mbn}=
\iota_{\mbm,\mbn}\,\Omega_{'\mbm,'\mbn}$
where
\begin{align}
'\mbm&=\cup_{m\in\mbm,m\neq -1}\{m+1\}\;\;\;\qquad\qquad\quad (1\in\mbn)\\
&=\cup_{m\in\mbm,m\neq -1}\{m+1\}\cup\{0\}\,\quad\qquad (1\notin\mbn)\\
'\mbn&=\cup_{n\in\mbn,n\neq 1}\{n-1\}\;\;\;\;\qquad\qquad\qquad (-1\in\mbm)\\
&=\cup_{n\in\mbn,n\neq 1}\{n-1\}\cup\{0\}\,\;\quad\quad\qquad (-1\notin\mbm)\,,
\end{align}
as can readily be verified. In the same way
this corresponds to $\bfG^{-1}$, formulae for which
appear in \cite[\S3.1]{stu12}. All together this determines
a unitary transformation $$\bfG\,\sum\psi_{\mbm,\mbn}\Omega_{\mbm,\mbn}
=\sum\iota_{\mbm,\mbn}\,\psi_{\mbm,\mbn}\,\Omega_{\mbm',\mbn'}$$ of $\mcH_0$
which induces \eqref{gt3}-\eqref{gt2}.
\qed

The transformation $\bfG$ commutes with $Q$ and so preserves $\mcH_0$, but it
does not commute with $Q^5$: for example 
$b_3^\dagger c_2^\dagger b_0^\dagger c_{1}^\dagger{\Omega_0}$ is mapped
into $b_2^\dagger c_3^\dagger c_2^\dagger b_{-1}^\dagger{\Omega_0}$, with the 
eigenvalue of $Q^5$ reducing by 2. 
Formally $Q^5\bfG^{-1}=\bfG^{-1}(Q^5-2)$ on $\mcF_0$.
The interpretation
of all these formulae is that {\em large gauge transformations can create and 
annihlate fermion/anti-fermion pairs} in a way which seems naively
to change the axial charge: an {\em anomaly}.
Nevertheless we have:
\begin{lemma}\la{gi}
The Schwinger 
regularizations of the axial charge \eqref{q5r}, and of the Hamiltonian 
\eqref{finh}-\eqref{coul}, are unchanged by the action of $\cm$.
\end{lemma}
\proof
This is straightfoward to check, see \cite[\S 3.4]{stu12}.
\qed

Now the gauge transformation $g_1$ acts on the connection as
$a\to a+\frac{2\pi}{L}$, and hence the requirement of gauge invariance
means that we should regard the Hamiltonian $H$ as an unbounded operator
defined on $\mcK$ with the following boundary conditions of 
{\em twisted periodicity}:
\beq\la{bc}
\Psi(\frac{2\pi}{L})=
\bfG^{-1}\Psi(0)
\quad\hbox{and}\quad
{\Psi'}(\frac{2\pi}{L})=
\bfG^{-1}{\Psi'}(0)
\,.
\eeq
(writing  prime for $\frac{d}{da}$). A suitable dense domain for
the Hamiltonian is $\mcD$, the space of smooth functions taking
values in $\mcF_0$ which satisfy this twisted periodicity condition,
i.e. the restriction to $[0,\frac{2\pi}{L}]$ of the
smooth $\mcF_0$-valued functions which satisfy $\Psi(a+\frac{2\pi}{L})
=\bfG^{-1}\Psi(a)$ for all $a\in\R$.
\begin{lemma}\la{dense} 
$\mcD\subset\mcK$ is dense in the norm $\|\,\cdot\,\|_{\mcK}$ on $\mcK$. 
The integration by parts formula $\langle\Psi',\Phi\rangle_{\mcK}=
-\langle\Psi,\Phi'\rangle_{\mcK}$ holds for $\Psi,\Phi$ in $\mcD$.
\end{lemma}
\proof
It suffices to first approximate $\Psi\in\mcK$ by simple functions
$\sum\id_{I_j}(a)f_j$ where $f_j\in\mcF_0$ and $I_j$ are measurable 
sets contained in a closed sub-interval of
$(0,\frac{2\pi}{L})$. Then approximate the characteristic functions
$\id_{I_j}(a)$ by smooth functions compactly supported in $(0,\frac{2\pi}{L})$.
The twisted periodicity condition is then trivially satisfied. The integration
by parts formula holds since $\bfG$ is unitary on $\mcH_0$.
\qed
\begin{remark}\la{vv}
The Fock vacuum $\Omega_0$, thought of as an element of 
$\mcK$ which is independent of $a$, does not satisfy \eqref{bc} and
is not gauge invariant. It follows that the interacting (or physical) 
vacuum cannot be proportional to $\Omega_0$, or indeed any of the unexcited
states $\Omega_P$, since $\cm$ maps these states into one another,
thus destabilizing the Fock vacuum.
We shall see in \S\ref{vacsec} that the physical vacuum is a linear combination
of states of the form $f_P(a)\Omega_P$.
\end{remark}
\begin{remark}\la{thet}
If $\Psi$ satisfies \eqref{bc} then $\tilde\Psi=e^{i\theta Q^{5,reg}}\Psi$
satisfies $\Psi(\frac{2\pi}{L})=e^{-2i\theta}\bfG^{-1}\Psi(0)$, corresponding
to a phase change in the definition of $\bfG$ (which is clearly allowed
by the above discussion). The parameter is called the $\theta$ parameter,
and this transformation shows that it does not give any new physics, but
rather corresponds to the choice of an equivalent representation for
$E^{tr}$ - see \cite[\S2 and \S6]{mant}.
\end{remark}
\section{Bosonization}
\setcounter{equation}{0}
\la{boso}

From \cite{liebmattis}, and earlier references therein, 
it is known that associated to a free massless fermionic field 
is a free real scalar field $\Phi$ with
conjugate momentum $\Pi$, given at fixed time by:
\begin{align}\la{dsf}
\begin{split}
\Phi(x)&=\sum \Phi_m e^{ik_m x}\,,\quad\Phi_m^\dagger=\Phi_{-m}\\
\Pi(x)&=\sum \Pi_m e^{ik_m x}\,,\quad\Pi_m^\dagger=\Pi_{-m}
\end{split}
\end{align}
where $k_m=2m\pi/L$ for $m\in\Z$, and with 
$[\Pi_{-m},\Phi_{m'}]=-\frac{i}{L}\delta_{mm'}$ 
(all other commutators being zero). This implies
the relation 
$[\Pi(x'),\Phi(x)]=-\frac{i}{L}\sum e^{ik_m(x-x')}=-i\delta(x-x')\,.$
The relevant representation of these commutation relations is related
to the particle structure and as such will be determined later. The
final conclusion will be the identification, after normal ordering and a shift
of the vacuum energy, of the Hamiltonian for the Schwinger
model with the Hamiltonian:
\beq\la{hss}
H_S=\,\frac{1}{2}\,\int_0^L\,\Bigl
(\Pi(x)^2+\partial \Phi(x)^2+\frac{e^2}{\pi}\Phi(x)^2\Bigr)\,dx
\,,
\eeq
for a real scalar field with mass $e/\sqrt{\pi}$. This is to be expected from
Schwinger's work \cite{sch62}.

The relation of $\Phi,\Pi$ to the fermionic and electromagnetic
fields of the Schwinger model is given by the formulae:
\begin{align}
\la{dfm}
\begin{split}
\Phi_m & = -\frac{\sqrt{\pi}}{ik_m}\jmath^0(m)\,,\qquad
\Pi_m ={\sqrt{\pi}}\jmath^1(m)\,,\qquad (m\neq 0)
\\
\Phi_0 & =  \frac{\sqrt{\pi}}{e^2}E^{tr}\,,\qquad\qquad
\Pi_0  =  \frac{\sqrt{\pi}}{L}Q^{5,reg}
=  \frac{\sqrt{\pi}}{L}(Q^{5}-\frac{aL}{\pi}-1)\,.
\end{split}
\end{align}
Here $\jmath^\mu(m)$ are the fourier modes of the current operators
$j^\mu=\overline\psi\gamma^\mu\psi$:
\beq
j^0=\sum\jmath^0(m)e^{ik_mx}\,,\qquad
j^1=
\sum\jmath^1(m)e^{ik_mx}\,.
\eeq
The $\jmath^\mu(m)$ may be obtained from the
fermionic creation/annihlation operators from the formulae (for $m\in\N$):
\beq\la{jri}
{\jmath^1}(m)=\frac{1}{L}\bigl[\varrho^R(-m)-\varrho^L(-m)\bigr]\,,\qquad
{\jmath^1}(-m)=\frac{1}{L}\bigl[\varrho^R(m)-\varrho^L(m)\bigr]\,,
\eeq
\beq\la{jr0}
{\jmath^0}(m)=\frac{1}{L}\bigl[\varrho^R(-m)+\varrho^L(-m)\bigr]\,,\qquad
{\jmath^0}(-m)=\frac{1}{L}\bigl[\varrho^R(m)+\varrho^L(m)\bigr]\,,
\eeq
where
\begin{align}\la{dd1}
\varrho^R(m)&=-\sum_{k>m}  c_k^\dagger c_{k-m}
+\sum_{0<k\leq m} b_{-k+m}^\dagger c_k^\dagger
+\sum_{k\geq 0} b_{k+m}^\dagger b_k\,,\\
\la{dd2}
\varrho^R(-m)&=-\sum_{k>m} c_{k-m}^\dagger c_{k}
+\sum_{0<k\leq m}c_k b_{-k+m}
+\sum_{k\geq 0} b_{k}^\dagger b_{k+m}\,,\\
\la{dd3}
\varrho^L(m)&=\sum_{k<-m} b_{k+m}^\dagger b_{k}
+\sum_{-m\leq k<0}c_{-k-m}b_{k}
-\sum_{k\leq 0}  c_{k}^\dagger c_{k-m}\,,\\
\la{dd4}
\varrho^L(-m)&=\sum_{k<-m} b_k^\dagger b_{k+m}
+\sum_{-m\leq k<0}b_k^\dagger c_{-k-m}^\dagger
-\sum_{k\leq 0} c_{k-m}^\dagger c_k \,.
\end{align}
The expressions \eqref{dd1}-\eqref{dd4} arise by considering
the fourier expansion of the density operators 
$\rho^{R,L}(x)=\frac{1}{2}\psi^\dagger(1\pm\gamma^5)\psi(x)$.
They are
precisely the quantities appearing in \cite[(3.1)-(3.4)]{liebmattis}
except for some conventions (ordering and the sign of the integer index 
on the $c_k$ operators has
been reversed.) Notice the following two important features:
\begin{itemize}
\item
Gauge invariance: $\bfG \rho^{R,L}(m)\bfG^{-1}=\rho^{R,L}(m)\,$;
\item
on the finite particle subspace $\mcF_0$ the
expressions \eqref{dd1}-\eqref{dd4} 
reduce to finite sums, and direct computation yields:
\beq\la{dc}
[\varrho^R(-m'),\varrho^R(m)]=
[\varrho^L(m'),\varrho^L(-m)]
=m\delta_{mm'}\,,
\eeq
for positive integral $m,m'$, other commutators being zero.
\end{itemize}
The relations \eqref{dc} imply the commutation relation 
$[\Pi_{-m},\Phi_{m'}]=-\frac{i}{L}\delta_{mm'}$ 
required to ensure that the field defined as
in \eqref{dsf} is a canonical scalar field.
The commutation relations
\eqref{dc} suggest that for $m\in\N$ the 
$m^{-\frac{1}{2}}\varrho^R(-m),m^{-\frac{1}{2}}\varrho^L(m)$ 
(resp. $m^{-\frac{1}{2}}\varrho^R(m),m^{-\frac{1}{2}}\varrho^L(-m)$)
represent annihlation (resp. creation) operators. In addition
notice that in \eqref{dd1} and \eqref{dd4} (resp. \eqref{dd2} and
\eqref{dd3}) the first and third
terms merely shift the momentum of fermions already present, while
the middle terms create (resp. annihlate) two fermions of opposite
chirality, so that $Q^5=2P$ is unchanged. This motivates the following result:

\begin{prop}[\cite{uhl67}, \S 4]\la{bos}
The subspaces $\mcH_0^P\subset\mcH_0$ are irreducible cyclic subspaces
for the algebra of operators generated by 
$\{\varrho^R(m'),\varrho^L(m)\}_{m,m'\in\Z-\{0\}}$, giving
rise to a Fock representation of the canonical commutation relations
with cyclic vector ${\Omega_P}$ which verifies
$\varrho^R(-m){\Omega_P}=0=\varrho^L(+m){\Omega_P}$ for $m\in\N$.
\end{prop}

In \cite{liebmattis}
it is pointed out that the fermionic
kinetic energy operator $$
H^0_D\,=\,\sum_{m\in\Z}\,|k_m|\,({b}^\dagger_{m}b_{m}+
c_m^\dagger c_{m})$$ 
has the same
commutation relations with the operators
$\{\varrho^R(m'),\varrho^L(m)\}_{m,m'\in\Z-\{0\}}$
as the nonnegative operator 
\beq\la{deft}
T=\frac{2\pi}{L}\sum_{m\in\N}\bigl(
\rho^R(m)\rho^R(-m)+\rho^L(-m)\rho^L(m)
\bigr)\,\geq\, 0
\eeq
(on each finite particle subspace $\mcF_0^P$). From this it follows from 
the above proposition that on $\mcF_0^P$
\beq\la{ki}
H^0_D\,=\,T\,+<P\;|H_D^0|\;P>=\,T\,+\frac{\pi}{2L}Q^5(Q^5-{2})
\eeq
(Kronig's identity). In fact the identity \eqref{ki} extends to an equality 
between self-adjoint operators on $\mcH_0$, see \cite[\S 5]{uhl67}.
Combining with \eqref{finh} we obtain the following bosonized
formula for \eqref{finh}:
\beq\la{finh2}
H_0=-\frac{e^2}{2L}\frac{d^2}{da^2}+\frac{\pi}{2L}(Q^{5,reg})^2
+T\,.
\eeq

\section{The vacuum for $H_0$}
\setcounter{equation}{0}
\la{vacsec}
Since integration by parts
is allowed by lemma \ref{dense} and
$T\geq 0$ as an operator inequality:
\begin{align}
\langle\,\Psi\,,H_0\,\Psi\,\rangle_{\mcK}
\geq\,\frac{e^2}{2L}\|\Psi'\|_{\mcK}^2\,+\,\frac{\pi}{2L}
\|Q^{5,reg}\Psi\|_{\mcK}^2\,,
\end{align}
on $\mcD$. Next, using the orthogonal decomposition
$\mcH_0=\oplus\mcH_0^P$, we can write 
$\Psi=\sum\Psi_P$ with $\Psi_P(a)\in\mcH_0^P\cap\mcF_0\,\forall a$ and 
since $Q^{5,reg}=2P-\frac{aL}{\pi}-1$ on $\mcH_0^P$ we get
\begin{align}
\langle\,\Psi\,,H_0\,\Psi\,\rangle_{\mcK}
\geq&\,\sum_P\Bigl[
\frac{e^2}{2L}\|\Psi_P'\|_{\mcK}^2\,+\,\frac{\pi}{2L}
\|(2P-\frac{aL}{\pi}-1))\Psi_P\|_{\mcK}^2
\Bigr]\,,\\
\geq&\,\sum_P\frac{e}{2\sqrt{\pi}}\|\Psi_P\|_{\mcK}^2
=E_0\,\|\Psi\|_{\mcK}^2\,,\la{lb}
\end{align}
where $E_0=\frac{e}{2\sqrt{\pi}}$,
using the standard lower bound for the oscillator which follows
from the commutation relation $[\frac{d}{da},a]=1$.
Thus $H_0\geq E_0$ on $\mcD$.

We will now show that
this lower bound $E_0$ is realized on states of the form
\beq
\Psi_0(a)=\sum f_P(a){\Omega_P}\,.
\eeq
The twisted periodic boundary condition \eqref{bc} translates into the 
requirements 
\beq\la{mc}
f_{P+1}(\frac{2\pi}{L})=f_P(0)\,,\quad\hbox{and}
\quad f'_{P+1}(\frac{2\pi}{L})=f'_P(0)
\eeq
for the
sequence of functions.
We aim to solve $H_S\Psi_0=E_0\Psi_0$ under these conditions.
Defining 
$$f(\tilde a)=f_P(\tilde a+\frac{2\pi P}{L}-\frac{\pi }{L})
\quad\hbox{for}\quad \tilde a\in I_P=
[-\frac{2\pi }{L}(P-\frac{1}{2})
\,,\,
-\frac{2\pi }{L}(P-\frac{3}{2})]
$$ 
gives
a function on the real line, and the eigenvalue equation 
$H_S\Psi_0=E_0\Psi_0$ is equivalent to the oscillator Schrodinger equation
\beq
-\frac{e^2}{2L}\frac{d^2 f}{d\tilde a^2}+\frac{L}{2\pi}\tilde a^2 f=E_0 f
\eeq
which has a solution for $E_0=e/(2\sqrt{\pi})$ proportional to
$e^{\frac{-L\tilde a^2}{2\sqrt{\pi} e}}$. Normalizing we define
$$f(\tilde a)=\,
\frac{L^{\frac{1}{4}}}{\pi^{\frac{3}{8}} e^{\frac{1}{4}}}\,
e^{\frac{-L\tilde a^2}{2\sqrt{\pi} e}}$$
and the vacuum state for $H_0$ is
\beq
\Psi_0(a)=\sum_{P\in\Z} f(a-\frac{2\pi }{L}(P-\frac{1}{2}))\Omega_P\,,
\la{vac0}
\eeq
with normalization $\|\Psi_0\|_{\mcK}=1$.
This recovers one of the results of
\cite{mant} but transferred to the positive energy representation.

\section{The coulomb interaction}\la{sec-coul}
\setcounter{equation}{0}
Next consider the Coulomb interaction, which is formally
$\frac{1}{2}e^2L\sum_{m\neq 0}{\jmath^0(-m)}k_m^{-2}{\jmath^0(m)}$. In order
to obtain a densely defined operator it is necessary to subtract off the
expectation with respect to the vacuum ${\Psi_0}$ in \eqref{vac0}. Thus we 
consider the quadratic form
$$
\frac{1}{2}e^2L\sum_{m\neq 0}\frac{1}{k_m^2}\bigl(
\|{\jmath^0(m)}\Psi\|_{\mcK}^2
-\|{\jmath^0(m)}{\Psi_0}\|_{\mcK}^2
\bigr).
$$
Noting that $\|\jmath^0(m)\Omega_P\|^2=|m|$ for each $P$ it is
easy to check from the exact expression \eqref{vac0} that this 
subtraction is equivalent to normal ordering with respect to the bosonic
algebra described in proposition \ref{bos}, and the corresponding
operator is
\beq
:H_{coul}:=\frac{1}{2}e^2L\sum_{m\neq 0}\frac{1}{k_m^2}
:\jmath^0(-m)\jmath^0(m):
\eeq 
with $:\jmath^0(-m)\jmath^0(m):$ given by
\beq
\frac{1}{L^2}\bigl(
\rho^R(m)\rho^R(-m)+\rho^L(m)\rho^R(-m)+\rho^R(m)\rho^L(-m)
+\rho^L(-m)\rho^L(m)
\bigr)
\eeq
for $m\in\N\,$, and
\beq
\frac{1}{L^2}\bigl(
\rho^R(-m)\rho^R(m)+\rho^L(m)\rho^R(-m)+\rho^R(m)\rho^L(-m)
+\rho^L(m)\rho^L(-m)
\bigr)
\eeq
for $-m\in\N\,$.
These two formulae can be combined into the formula
\beq
:\jmath^0(-m)\jmath^0(m):\,=\,
\frac{1}{L^2}\bigl(
\rho^R(m)\rho^R(-m)+\rho^L(m)\rho^R(-m)+\rho^R(m)\rho^L(-m)
+\rho^L(m)\rho^L(-m)-|m|\bigr)\,.
\nonumber\eeq
The fact that this operator is densely defined is now straightforward
since up to a factor $\sqrt{|m|}$ the $\rho^{L,R}$ are just 
creation/annihlation operators, and the above expressions are normal ordered
and are densely defined and essentially self-adjoint on the domain $\mcF_0$.
(See \cite[Theorem 1]{dimock} for a dense domain 
for the case of massive fermions when bosonization is not available
in the simple form used here.)

Referring to \eqref{dfm} it is apparent that $:H_{coul}:=
\frac{e^2L}{2\pi}\sum_{m\neq 0}
:\Phi^\dagger_{m}\Phi_{m}:\,\,
,$
and it is straightforward to check 
that formally 
$$H'\,=\,T+:H_{coul}:
\,=\,\frac{L}{2}\,\sum_{m\neq 0}\,\Bigl(\bigl(k_m^2+\frac{e^2}{\pi}\bigr)\Phi_m^\dagger\Phi_m
+\Pi_m^\dagger\Pi_m\Bigr)\,.
$$
Note that $T$ is normal ordered by definition, see \eqref{deft}.
However $\Phi$ is not yet in the right representation:
to complete the identification of the full Hamiltonian 
in \eqref{finh2}-\eqref{q5r} with \eqref{hss}, the
Hamiltonian of a mass $\mu=e/\sqrt{\pi}$ scalar field,
it is sufficient to obtain a unitary Bogoliubov transformation
which diagonalizes $T+:H_{coul}:$ and puts the fields
into the appropriate positive energy representation. 
Following \cite{liebmattis,uhl67} this is done by means of a 
Bogoliubov transformation as follows: let
${V_m}=k_m^{-2}$ and $\lambda=\frac{1}{2}e^2$, then defining
the real-valued even function $\zeta(m)$ by
$$\frac{\lambda V_m}{\lambda V_m+\pi}=-\tanh 2\zeta(m)$$
the self-adjoint operator
$$
Z=\frac{2\pi i}{L}\sum_{m\neq 0}\frac{\zeta(m)}{k_m}\rho^R(m)\rho^L(-m)
$$
generates a unitary transformation $\bfU=e^{iZ}$ with the property that
(putting transformed operators into boldface):
\begin{align}\la{ff}
\bfH'=\bfU H'\bfU^{-1}\,&=\,\frac{2{\pi}}{L}\sum_{m\in\N}
\bigl(1+\frac{2\lambda V_m}{\pi}
\bigr)^\frac{1}{2}\bigl(\rho^R(m)\rho^R(-m)+\rho^L(-m)\rho^L(m)\bigr)\\
&\qquad\quad+\,\frac{2\pi}{L}\sum_{m\in\N}
\, m\Bigl[\bigl(1+\frac{2\lambda V_m}{\pi}
\bigr)^{\frac{1}{2}}
-\frac{\lambda V_m}{\pi}-1\Bigr]\,.\nonumber
\end{align}
For $m\in\N$ define 
\beq
A_m^\dagger=-im^{-\frac{1}{2}}\rho^R(m)\,,\qquad
A_m=im^{-\frac{1}{2}}\rho^R(-m)\,,\eeq
and
\beq
A_{-m}^\dagger=im^{-\frac{1}{2}}\rho^L(-m)\,,\qquad
A_{-m}=-im^{-\frac{1}{2}}\rho^L(m)\,,\eeq
so that $[A_m,A_{m'}^\dagger]=\delta_{mm'}$ for a non-zero
integral $m,m'$. In terms of these operators
\begin{align}
\bfH'\,=\,
\sum_{m\in\N}\bigl(k_m^2+\frac{e^2}{\pi}\bigr)^{\frac{1}{2}}
\Bigl[A_m^\dagger A_m+A_{-m}^\dagger A_{-m}\Bigr]\,+\,C_0
\end{align}
where $C_0\in (\,-\infty\,,\, 0\,)$ is the second line of \eqref{ff}.

This allows an identification of the interacting vacuum for the
full Hamiltonian as follows. The Hamiltonian is given by
\begin{align}\la{fham}
\bfH\,=\,\bfU H\bfU^{-1}\,=\,
-\frac{e^2}{2L}\frac{d^2}{da^2}+\frac{\pi}{2L}(Q^{5,reg})^2+
\sum_{m\in\N}\bigl(k_m^2+\frac{e^2}{\pi}\bigr)^{\frac{1}{2}}
\Bigl[A_m^\dagger A_m+A_{-m}^\dagger A_{-m}\Bigr]\,
+C_0\,.
\end{align}
We use boldface to distinguish the dressed forms of the
various states, thus defining $\bfOm_P=\bfU^{-1}\Omega_P$ we see that
the vacuum energy is $E_0+C_0<E_0$ with corresponding eigenfunction
\beq\la{iv}
\bfp(a)=\sum_{P\in\Z} f(a-\frac{2\pi P}{L})\bfOm_P\,,
\eeq
The effect of the Coulomb term on the vacuum is to shift the
energy down by a finite amount $C_0<0$ and to map the
non-interacting unexcited states $\Omega_P$ to their dressed
versions $\bfOm_P$.

The fourier components of the dressed scalar field are given by
\begin{align}\la{fcdf}
\bfPhi_m\,&=\,\frac{1}{\sqrt{2\omega_m L}}(A_m+A_{-m}^\dagger)\,,\\
\bfPi_m\,&=\,-i\sqrt{\frac{{\omega_m}}{2L}}(A_m-A_{-m}^\dagger)\,,
\end{align}
with $\omega_m=\sqrt{k_m^2+\frac{e^2}{\pi}}$ for $m\neq 0$, in terms of
which 
\begin{align}
\bfH\,&=\,\bfU H\bfU^{-1}\,=\,
-\frac{e^2}{2L}\frac{d^2}{da^2}+\frac{\pi}{2L}(Q^{5,reg})^2+
\,\frac{L}{2}\,\sum_{m\neq 0}\,:\bigl(\bfPi_m^\dagger\bfPi_m+(k_m^2+\frac{e^2}{\pi})
\bfPhi_m^\dagger\bfPhi_m\bigr):\,
\,
+C_0\notag\\
\,&=\,\frac{L}{2}\,\sum_{m\in\Z}\,:\bigl(\bfPi_m^\dagger\bfPi_m+(k_m^2+\frac{e^2}{\pi})
\bfPhi_m^\dagger\bfPhi_m\bigr):
\,
+C_0\,.\la{fdham}
\end{align}
(This formula is the normal ordered version of \eqref{hss}. 
The action of the dressing transformation $\bfU$ is trivial on the
$m=0$ components, so that $
\bfPhi_0=\Phi_0 =  \frac{\sqrt{\pi}}{e^2}E^{tr}\,,$ and 
$\bfPi_0=\Pi_0  =  \frac{\sqrt{\pi}}{L}Q^{5,reg}$.) The existence of
a self-adjoint extension is now straightforward but nevertheless it is
interesting to see the role of twisted periodicity in ensuring that
$\mcD$ is a domain of essential self-adjointness in the proof of the
following theorem.

\begin{theorem}\la{sadj}
The
symmetric operator $\bfH$ is essentially
self-adjoint on $\mcD\subset\mcK$ (with self-adjoint 
extension also written $\bfH$).
\end{theorem}
\proof
Since $\bfH$ is bounded below it is sufficient to show that
$\bfH+\lambda_0$ has dense range for large $\lambda_0>0$ by 
\cite[theorem X.26]{MR751959}. 
Consider the orthonormal set of vectors in $\mcH_0^P$ of the form
\beq
\Omega_P^{\bf n}\,=\,const.\, \prod_{m\neq 0}(A_{m}^\dagger)^{n_m}\Omega_P
\eeq
labelled by ${\bf n}=\{n_m\}_{m\in\Z-\{0\}}\,,$ with $n_m\in\N\cup\{0\}$ and
only a finite number of the $n_m$ nonzero. Since 
$\bfG \rho^{R,L}(m)\bfG^{-1}=\rho^{R,L}(m)\,$ it follows from
lemma \ref{ag} that 
\beq
\bfG\Omega_P^{\bf n}\,=\,\Omega_{P-1}^{\bf n}\,.
\eeq
(Strictly speaking to achieve this it is necessary specify that
the normalization constants in the definition of $\Omega_P^{\bf n}$
above are chosen independent of $P$, i.e. without any additional
$P$ dependent phase factors.)
Now linear combinations of
functions of the orthornomal set 
$e^{i l L a}\Omega_{P_0}^{\bf n}$ span a dense
set in $\mcK$ and so  as first stage we want to solve
$(\bfH+\lambda_0)\Psi=e^{i l L a}\Omega_{P_0}^{\bf n}$ 
for $\Psi\in\mcD$. The boundary condition
means that the solution must involve all values of $P$ - 
as in \eqref{vac0} we obtain a solution of the form 
$\Psi=\sum_Pf_P(a)\Omega_P^{\bf n}$.
We can generate such a solution as follows: write $\R=\cup I_P$ where
$I_P=[-\frac{2\pi}{L}(P-\frac{1}{2}),-\frac{2\pi}{L}(P-\frac{3}{2})]$ and
define $f_P(a)=f(a-\frac{2\pi }{L}(P-\frac{1}{2}))\,,0\leq a\leq \frac{2\pi}{L}
\,,P\in\Z\,,$ 
where $f(\tilde a)$
is a function on $\R$ which solves
$$
-\frac{e^2}{2L}\frac{d^2f}{d{\tilde a}^2}+\frac{L}{2\pi}{\tilde a}^2f
+(E^{ex}_{\bf n}+\lambda_0+C_0)f=e^{i l L {\tilde a}-il\pi}
\id_{I_{P_0}}(\tilde a)\,,
$$
and $E^{ex}_{\bf n}=\sum\omega_m n_m$ is the bosonic excitation energy. 
(The twisted boundary conditions \eqref{bc} imply the matching conditions
for $f,f'$ exactly as in \eqref{mc}).
For $\lambda_0+C_0\geq 0$ it
follows as in \eqref{lb} that $\|f\|_{L^2(\smr)}\leq const.
=|I_{P_0}|/E_0=2\pi/(LE_0)$. This
is independent of $P,P_0,l,\mathbf{n}$,
and so since
the $\Omega_P^{\bf n}$ are of unit length in Fock space 
and orthogonal for different $P$ we get
$\|\Psi\|_{\mcK}^2=\|\sum_Pf_P(a)\Omega_P^{\bf n}\|^2_{\mcK}\leq 
\sum_P\|f_P(a)\|^2_{L^2([0,\frac{2\pi}{L}])}=\|f\|^2_{L^2(\smr)}\leq (const.)^2$.
Finally, by linearity 
$(\bfH+\lambda_0)\Psi
=\sum c_{l,\mathbf{n},P_0}e^{i l L a}\Omega_{P_0}^{\bf n}$ 
has a solution in $\mcD$ satisfying $\|\Psi\|_{\mcK}^2
\leq (const.)^2\sum |c_{l,\mathbf{n},P_0}|^2
\leq (const.)^2\|\sum c_{l,\mathbf{n},P_0}e^{i l L a}\Omega_{P_0}^{\bf n}\|_{\mcK}^2$, and
since finite linear combinations of 
the $e^{i l L a}\Omega_{P_0}^{\bf n}$
constitute a dense set this
proves density of the range of $\bfH+\lambda_0$. 
\qed

\section{The anomaly equation}\la{anom}
\setcounter{equation}{0}
In this section we derive the anomaly equation \eqref{ae},
which is the
quantum analogue of the classical conservation law
$\partial_\mu j^{5,\mu}_{classical}=0$ discussed following \eqref{ceom}.
In fact we will work with the dressed fields, indicated by the
use of boldface as in \S\ref{sec-coul}, and we will derive the
fourier transformed version of \eqref{ae}:
\beq\la{fae}
\partial_t\boldsymbol{\jmath^1}(m)+ik_m\boldsymbol{\jmath^0}(m)=-\frac{1}{\pi}
{{\mathbf{E}}}_m
\eeq
where \beq
{\mathbf{E}}=\sum_{m\in\Z}{\mathbf{E}}_me^{ik_m x}={\mathbf{E}}^{tr}+{\mathbf{E}}^{long}
={\mathbf{E}}^{tr}+\sum_{m\neq 0} {\mathbf{E}}^{long}_m e^{ik_m x}\,,
\eeq 
where, referring to \S\ref{secint}, the operator corresponding
to the longitudinal electric field has
fourier components ${\mathbf{E}}^{long}_m=-e^2\boldsymbol{\jmath^0}(m)/(ik_m)$,
as determined by the Gauss law. Therefore by \eqref{dfm}
${\mathbf{E}}_m=e^2\bfPhi_m/\sqrt{\pi}$. (As remarked at the end of 
\S\ref{sec-coul} the dressing 
operation acts trivially for the case $m=0$,
and so $\mathbf{E}_0=E_0=E^{tr}$ and $\bfPhi_0=\Phi_0$ etc.)
The anomaly equation \eqref{fae}
is therefore a consequence of 
(or, indeed, essentially equivalent to) the Heisenberg equation of motion
$
\dot\bfPi_m\,=\,-\omega_m^2\,\bfPhi_m$
since, by the definitions \eqref{dfm}:
\begin{align}
\partial_t\boldsymbol{\jmath^1}(m)\,+ik_m\boldsymbol{\jmath^0}(m)
=\frac{1}{\sqrt{\pi}}(\dot\bfPi_m+k_m^2\bfPhi_m)
=\frac{1}{\sqrt{\pi}}(-\frac{e^2}{\pi}\bfPhi_m)
=-\frac{1}{{\pi}}{\mathbf{E}}_m\,.
\end{align}
This holds as an equality of operator valued distributions
on $\R$.

We can see now that the anomaly occurs because we have enforced gauge
invariance through the definitions \eqref{finh}-\eqref{q5r} which 
were derived from Schwinger gauge invariant regularization. But as was
made explicit in \S\ref{mod} the action of large gauge transformations
on the  positive energy representation Fock space does not respect
the chiral phase invariance of the classical theory and $Q^5$ is not
invariant under this group action. Thus enforcing gauge invariance of the
Hamiltonian under this group action necessarily leads to a theory
which does not respect the chiral symmetry - it can be traced back to the
introduction of a non-invariant vacuum (``sea-level'') in the positive 
energy representation. It is interesting to compare this with the picture
of the spectral flow explained in \cite{mant} and \cite[\S6.6]{jack}, 
in which the anomaly
arises from the infinitely deep Dirac sea of filled levels 
(after regularization).

\section{Concluding remarks and relation with previous work.}
\setcounter{equation}{0}
Since \cite{sch62} 
there have been many treatments of the Schwinger model in the physics 
literature, see the bibliography in \cite{mant} or \cite{tsv,zj} for a 
textbook treatment. The physics of the model and its massive generalization
are discussed in \cite{cjs}.
Much of the previous rigorous mathematical work on two dimensional 
quantum electrodynamics, 
both the massless case of the Schwinger model studied here and the 
general massive case of $(QED)_2$, has employed functional integration
in Euclidean space-time - see \cite{f76,fs,ito2,wc79,w80}. These developments
give a rigorous treatment of the Schwinger model, though with less transparency
in respect of the points itemized below. 
In \cite{ito1} the Hamiltonian approach is pursued within the St\"uckelberg
indefinite metric formalism.
In \cite{dimock}
the Coulomb gauge Hamiltonian for $(QED)_2$ 
is shown to be densely defined and a comparison
with functional integral methods is made. However it should be said that
the problem studied in \cite[\S2]{dimock}, restricted to the massless case,
is strictly speaking a different model to that considered here: it is
Coulomb gauge $(QED)_2$ on $\R$ restricted by periodic 
boundary conditions to the circle, rather than $(QED)_2$ on the circle put
into Coulomb gauge.
(It is assumed that it is possible to eliminate the spatial component of the
potential completely, which is not possible on the circle with a periodic gauge 
transformation but {\em is} possible with the whole real line
as spatial domain.
As a consequence the modular group $\cm$ does not appear in 
\cite{dimock}, and there is no description of the anomaly.)

In this paper we have provided a Hamiltonian formulation of the Schwinger model
on the circle, incorporating the insights from \cite{mant} (regarding
the modular group and the anomaly) into
the positive energy representation. The advantage of this
representation is that a rigorous
definition of the Hamiltonian  $H=H_0+:H_{coul}:$ from \eqref{finh}-\eqref{coul}
as a self-adjoint operator is then easily achieved
through bosonization. Noteworthy conclusions are:
\begin{itemize}
\item There is a nontrivial action of $\cm=\Z$, 
the modular group of large gauge 
transformations, on the non-interacting Fock vacuum $\Omega_0$ given in 
\ref{mod} which maps $\Omega_0$ to the unexcited states $\Omega_P$. The
$\Omega_P$ are eigenstates of the (unregularized) axial charge operator 
$Q^5$ with
the following property: no excited fermionic state is occupied which 
has higher energy than an unoccupied one (amongst states with the 
correct sign of $Q^5$).
\item It is necessary to take into account this action of $\cm$
in the definition of the Hamiltonian and currents in order to 
obtain the ``correct'' gauge invariant expressions in \eqref{finh}-\eqref{q5r}.
These expressions, which are derived in \cite{stu12} using
Schwinger point-splitting regularization, contain terms which give 
rise to the anomaly in the chiral conservation law \eqref{ae}
and the mass of the 
fundamental boson described in \S\ref{boso}.
\item Even without ``turning on'' the Coulomb interaction the 
interaction between the fermions and the spatial component of the 
electromagnetic potential $a$ destabilizes the gauge variant non-interacting 
vacuum $\Omega_0$, producing an interacting vacuum which is gauge invariant 
(as expressed by the twisted periodicity condition \eqref{bc}).
There is an explicit formula for the interacting vacuum as a 
linear combination of segments of a gaussian tensored with $\Omega_P$, 
wrapped around the circle $S^1=\R/(2\pi/L)$ in such a way as to satisfy 
\eqref{bc} - see \eqref{vac0}.
\item The effect of turning on the Coulomb interaction is to transform
the unexcited states $\Omega_P$ into dressed versions $\bfOm_P$, in terms
of which the vacuum takes the same form - see \eqref{iv}.
\end{itemize}

It is to be hoped that the description of the Schwinger model and
its anomaly offered here will, by virtue of its transparency, be helpful
in further developments such as mass perturbation theory (\cite{cjs,fs}), 
curved space-time 
and generalization to non-abelian gauge groups.
\small
\section*{Acknowledgements}
\setcounter{equation}{0}
I would like to thank Nick Manton for many helpful discussions
regarding the Schwinger model and \cite{mant} in particular.

\end{document}